# The low-temperature highly correlated quantum phase in the charge-density-wave 1T-TaS$_2$ compound

Marie Kratochvilova[1,2], Adrian D. Hillier[3], Andrew R. Wildes[4], Lihai Wang[5], Sang-Wook Cheong[5], and Je-Geun Park[1,2]

A prototypical quasi-2D metallic compound, 1T-TaS$_2$ has been extensively studied due to an intricate interplay between a Mott-insulating ground state and a charge density-wave (CDW) order. In the low-temperature phase, 12 out of 13 Ta$^{4+}$ 5$d$-electrons form molecular orbitals in hexagonal star-of-David patterns, leaving one 5$d$-electron with $S = \frac{1}{2}$ spin free. This orphan quantum spin with a large spin-orbit interaction is expected to form a highly correlated phase of its own. And it is most likely that they will form some kind of a short-range order out of a strongly spin-orbit coupled Hilbert space. In order to investigate the low-temperature magnetic properties, we performed a series of measurements including neutron scattering and muon experiments. The obtained data clearly indicate the presence of the short-ranged phase and put the upper bound on ~ 0.4 $\mu_B$ for the size of the magnetic moment, consistent with the orphan-spin scenario.

## INTRODUCTION

The instability of charge density waves (CDW) found in low-dimensional electron systems of layered materials has attracted enormous attention recently. 1T-TaS$_2$ is a prototypical quasi-2D metallic compound with a strong electron-phonon coupling responsible often for CDW instabilities. Upon cooling, it undergoes a series of first-order phase transitions to CDW, Mott and superconducting phases (see e.g. ref. 1 and references therein).

1T-TaS$_2$ bulk crystal has a lamellar structure, with each layer composed of a triangular lattice of Ta atoms: which is then sandwiched by S atoms in an octahedral TaS$_6$ coordination with a weak Van der Waals bonding between the layers. Above ~ 540 K, the structure is unmodulated trigonal with *P-3m1* symmetry; below which the triangular lattice exhibits a series of structural modulations[2, 3]. First, an incommensurate CDW phase sets in at $T$ ~ 540 K. Upon further cooling, the structure changes to a nearly commensurate phase at $T_{nCDW}$ ~ 350 K. Finally, the material turns into a Mott insulating phase with in-plane $\sqrt{13} \times \sqrt{13}$ superlattice distortion, which coexists with a commensurate CDW phase below $T_{CDW}$ ~ 180 K. Superconductivity emerges in 1T-TaS$_2$ below ~ 2 K by introducing disorders[4]. Recent angle-resolved photoemission spectroscopy experiments suggest that the melted Mott state and the superconductivity coexist in real space[5] providing better understanding of the interplay between electron correlation, charge order, and superconductivity. Unlike other Mott insulators, the CDW superlattices play the role of localization centers in the ground state of 1T-TaS$_2$.

[1]Center for Correlated Electron Systems, Institute for Basic Science, Seoul 08826, Korea; [2]Department of Physics and Astronomy, Seoul National University, Seoul 08826, Korea; [3]ISIS, STFC, Rutherford Appleton Laboratory, Didcot OX11 0QX, United Kingdom; [4]Institut Laue-Langevin, 156X, 38042 Grenoble Cedex, France; [5]Rutgers Center for Emergent Materials and Department of Physics and Astronomy, Rutgers University, Piscataway New Jersey 08854, USA

Correspondence: Marie Kratochvilova (kratm6bm@snu.ac.kr) or Je-Geun Park (jgpark10@snu.ac.kr)

According to the current understanding, the CDW phase is composed of molecular orbitals of 13 Ta atoms, forming a hexagonal pattern of so-called David-star clusters. As proposed more than three decades ago[6], each $Ta^{4+}$ of 1T-TaS$_2$ provides one 5$d$ electron, and thus there are 13 5$d$-electrons per each David star. Out of these 13 electrons, 12 electrons form 6 covalent bonds and so they become inactive both electrically and magnetically, leaving one 5$d$-electron localized well inside the David star. This orphan electron provides $S = ½$ per each David star, and becomes localized at low temperatures ($T < 60$ K) in the center of the David star[7, 8]. We note that in the 2D quantum triangular lattice systems, a combination of reduced dimensionality, geometric frustration, and a small spin value significantly enhances quantum fluctuations, which can easily induce exotic quantum states. In contrast to other $S = ½$ triangular systems showing rich phase diagrams such as Cu$RE_2$Ge$_2$O$_8$[9], Ba$_3$CoSb$_2$O$_9$[10], and others[11, 12], there is no apparent magnetic ordering in the insulating ground state of 1T-TaS$_2$ down to the lowest measured temperatures.

Instead, a commensurate CDW phase forms most likely through Mott-Hubbard physics with possibly intricate electron–phonon and electron-electron interactions[13]. This then leaves the orphan spin ($S = 1/2$) with a large spin-orbit interaction in the low-temperature CDW phase. How this orphan quantum spin behaves at low temperatures has been at the focus of current attention and hence extensively studied both theoretically[7, 8, 14] and experimentally[13, 15-17]. There is even a suggestion that it may form a highly correlated, "hidden" phase of its own and just recently[18], 1T-TaS$_2$ was suggested to be considered as a quantum spin liquid. Yet, none of the experimental works focused directly on the investigation of this low-temperature phase with a very recent exception[19]. It is thus interesting to ask whether some kind of a short-range ordering would be formed out of the strongly spin-orbit coupled Hilbert space, in which a magnetic moment $\mu = g_s\sqrt{(S(S+1))}/13$ $\mu_B \sim 0.13$ $\mu_B$ for $S = ½$ per each David star is expected.

In this work, we tried to address the question of a possible low-temperature correlated phase. For this, we have performed extensive bulk measurements, polarized neutron diffraction and $\mu$SR studies covering a wide range of time scales and spatial resolution. $\mu$SR and diffuse neutron scattering are ideal experimental techniques for investigating the weak magnetic signals and providing information on a correlation length of magnetic order, respectively.

**RESULTS**

We measured the magnetic susceptibility in the temperature range of 2 - 400 K with magnetic field applied along the hexagonal basal plane using several samples from different batches (see Fig. 1a and b). Diamagnetic signal dominates the data from high temperatures down to ~50 K, except for intermediate interruptions by two jumps at $T_{CDW}$ and $T_{nCDW}$, which signal a transition from the commensurate to the nearly commensurate phase and then to another incommensurate phase, respectively. While the transition to the incommensurate phase takes place at $T_{nCDW} \sim 355$ K, the transition at $T_{CDW}$ shows a remarkable hysteresis as shown in the inset of Fig. 1b and reaches values of ~ 220 K upon warming and ~ 160 K upon cooling. Interestingly, all the $\chi$(T) curves exhibit a strong paramagnetic Curie-Weiss tail below ~ 50 K, suggesting that some kind of a short-range correlation might exist at low temperatures. Magnetic impurities were excluded as a source of the low-temperature Curie contribution, considering the fact that we consistently observed a comparable low-temperature tail of almost similar magnitude for all the five samples prepared from raw materials of different purity. We also not that the Curie constant is too high to be explained purely by an extrinsic effect[20]. Our estimate of ~ 8 $\mu_B$/f.u. for the impurity moments is a too big a value if we adopt the impurity based explanation. All the known structure phases of the 1T-TaS$_2$ compound are summarized in Table 1.

**Table 1.** 1T-TaS$_2$ phase diagram

| Temperature | Phase |
|---|---|
| $T > 540$ K | Trigonal with *P-3m1* symmetry |
| 540 K $> T >$ 355 K | Incommensurate |
| 350 K $> T >$ 160 K | Nearly commensurate (upon cooling) |
| 160 K $> T$ | Commensurate Mott insulator |
| 60 K $> T$ | Localization of the free electron |

The magnetic susceptibility was fitted using the following equation with a diamagnetic/Pauli paramagnetic contribution $\chi_0$ and a Curie-type paramagnetic contribution in the temperature region from 2 to 200 K;

$$\chi = \frac{C}{T-\theta} + \chi_0, \qquad (1)$$

where $C$ is the Curie constant and $\theta$ is the paramagnetic temperature. The temperature dependence of the inverse magnetic susceptibility and the respective fits are shown in Fig. 1d. Parameters of the Curie-Weiss fit are also summarized in Table S1 in the Supplemental Material. The single crystals #1 and #2 are not diamagnetic down to ~ 50 K, probably due to a small presence of the 2H-polymorph in the samples (discussed in the Supplemental Material) which reveals itself by a tiny bump at ~ 75 K corresponding to the CDW transition[21] (marked by the arrow in Fig. 1b). We note that the size of the effective magnetic moment and the Curie-Weiss temperature ($\theta_{CW}$) is similar for all the measured samples with $\mu_{eff}$ ~ 0.08 $\mu_B$/f.u. and $\theta_{CW}$ ~ 0.02 K, respectively. Therefore, we observe ~ 50 % reduction of the observed $\mu_{eff}$ value from the expected spin-only value of ~ 0.13 $\mu_B$ for one magnetic moment per each David star. On the other hand, the small value of $\theta$ indicates weak magnetic correlations present at low temperatures.

In order to investigate the thermodynamic nature of the transitions, we also measured the specific heat. The specific heat collected upon cooling from 400 down to 0.7 K is shown in Fig. 1a in comparison to the magnetic susceptibility data. In accordance with the magnetic susceptibility, the specific heat reveals a typical 1$^{st}$-order transition from the incommensurate to the nearly commensurate phase at $T_{nCDW}$ ~ 355 K. However, we failed to observe any lambda-like peak signifying the 1$^{st}$-order transition into the commensurate phase in contrast to the behavior of $\chi$(T) measured on the identical sample. Instead, the $C_p/T$(T) curve shows a moderate decrease followed by a clear change of the slope at $T_{CDW}$ ~ 150 K. Similar discontinuous decrease/increase (depending on the rise/decrease of temperature) of the specific heat in the vicinity of $T_{CDW}$ was observed in the case of incommensurate-to-commensurate transition in 2H-TaSe$_2$[22]. Moreover, a discontinuity in the specific heat of 1T-TaS$_2$ was predicted by the calculation based on the Born von Karman model[23]. Upon further cooling, the $C_p/T$(T) curve reveals a sharp drop at ~ 50 K, which may be related to the upturn in the magnetic susceptibility at a similar temperature range.

The high-temperature specific heat data shown in Fig. 1c were modeled by the Debye equation. The phonon contribution can be written as

$$C_{ph} = 9NR \left(\frac{T}{\theta_D}\right)^3 \int_0^{\theta_D/T} \frac{x^4 e^x dx}{(e^x-1)^2}. \qquad (2)$$

Our analysis produces the following parameters for the best fit: the Debye temperature of $\theta_D$ ~ 450 K, the Sommerfeld coefficient of $\gamma = 1.84$ mJ.mol$^{-1}$.K$^{-2}$, and the coefficient $\beta = 0.31$ mJ.mol$^{-1}$.K$^{-4}$ as shown in the upper inset of Fig. 1c. We note that these values correspond rather well to those obtained from the previous experiment carried on powder sample[24]. Interestingly enough, the value of $\gamma$ is close to the theoretical value of 3.56 mJ.mol$^{-1}$.K$^{-2}$ from DFT calculations[25]. We note that $C_p$(T) shows a modest increase below ~ 1 K, indicating some kind of residual entropy still remaining at very low temperatures.

After subtracting off the estimated lattice and electron contributions, we can get a broad hump centered at ~ 50 K in the magnetic heat capacity $C_{mag}$, indicative of some kind of short-ranged order. Integrating the peak area yields an entropy released by this correlated phase, which is ≈ 40 % of the theoretical maximum entropy $R\ln(2)$ = 5.76 J.mol$^{-1}$ K$^{-1}$.

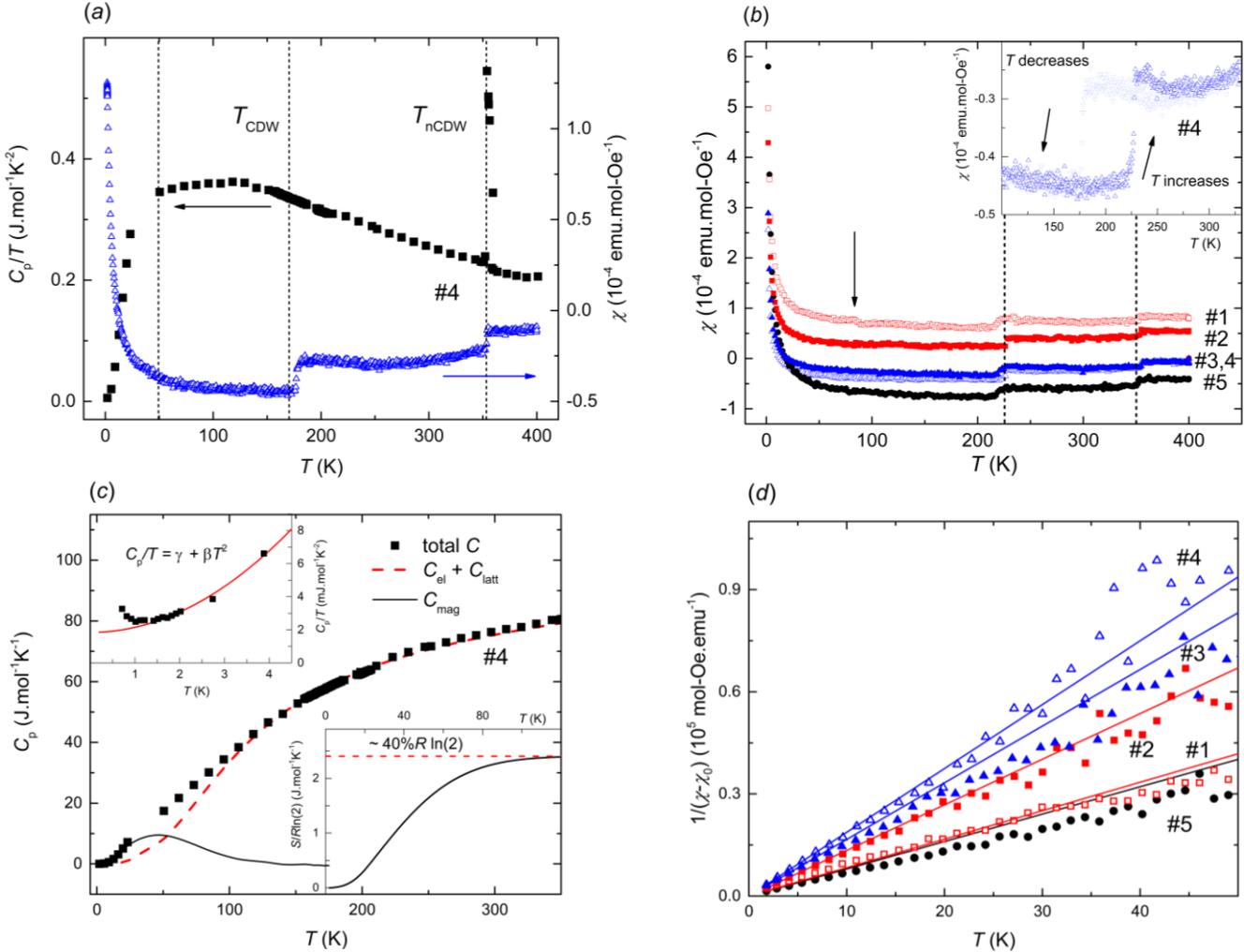

**Fig. 1** Magnetic and thermodynamic properties of 1T-TaS$_2$ **a** Comparison of the temperature dependence of specific heat and magnetic susceptibility of the 1T-TaS$_2$ single crystal #4 measured upon cooling. The dashed lines mark the position of the transitions at $T_{CDW}$ ~ 160 K and $T_{nCDW}$ ~ 355 K and the temperature ~ 50 K, below which an upturn in both data sets is observed. **b** The temperature dependence of the magnetic susceptibility measured on five 1T-TaS$_2$ single crystals in magnetic field of 0.2 T upon warming. The dashed lines mark the transition at $T_{CDW}$ ~ 210 K and $T_{nCDW}$ ~ 355 K. The inset shows the blown-up picture of the data near $T_{CDW}$ for the single crystal #4 measured upon cooling and warming. **c** Total specific heat of the 1T-TaS$_2$ single crystal #4 (closed squares) with the estimated lattice and electronic (dashed line) contribution and the magnetic (solid line) contribution. The low-temperature region of the specific heat is shown in the upper inset. The red line is the Sommerfeld fit to the data as given by the equation $C_p/T = \gamma + \beta T^2$. The magnetic entropy is shown in the lower inset. Approximately 40% of the theoretical maximum entropy (5.76 J.mol$^{-1}$ K$^{-1}$) is released as marked by the red dashed line. **d** The temperature dependence of the inverse magnetic susceptibility of the five 1T-TaS$_2$ single crystals from **b** and their respective Curie-Weiss fits

When taken together, the analysis of the Curie-Weiss fits, the low-temperature specific heat and the magnetic susceptibility provided clear evidence in support of the presence of a low-temperature correlated phase. In order to get a more direct insight into the low-temperature ground state of 1T-TaS$_2$, we performed neutron scattering and muon spin relaxation experiments. Fig. 2 shows the magnetic scattering data $(d\sigma/d\Omega)_{mag}$ of the 1T-TaS$_2$ compound. Although the signals are relatively weak, closer inspection of the data reveals a bump emerging in the $Q$–range from 0.7 to 1.2 Å$^{-1}$ (for more information about the signal quality, see the Supplemental Material). To emphasize the possible presence of a broad peak marked by the red arrow in the inset of Fig. 2, the original data were shifted upwards by a factor of 0.0075 barn.st$^{-1}$f.u.$^{-1}$.

We can model the data by the expression for the magnetic neutron-scattering differential cross section,

$$\left(\frac{d\sigma}{d\Omega}\right)_{mag} = \frac{2}{3}\left(\frac{\gamma_n r_0}{2}\right)^2 f^2(\vec{Q}) g_s^2 S(S+1) \times \left[1 + \sum_{n=1}^{\infty} Z_n \frac{\langle S_0 \cdot S_n \rangle}{S(S+1)} \frac{\sin QR_n}{QR_n}\right] - \left(\frac{d\sigma}{d\Omega}\right)_{bkg}, \quad (3)$$

which is valid in the case of a disordered magnetic configuration. The background term $\left(\frac{d\sigma}{d\Omega}\right)_{bkg}$ = 0.0075 barn.st$^{-1}$f.u.$^{-1}$ was added into the equation (3) to compensate the systematic errors which play an important role in our data with the low signal-to-noise ratio. Details are discussed in the Supplemental Material. Using the formula, we can calculate the average spin-spin correlation functions $\langle S_0 \cdot S_n \rangle$ for the $n^{th}$ nearest neighbor radial shell around a central Ta atom at the origin. Here, $\gamma_n$ is the neutron gyromagnetic ratio, $r_0$ is the classical electron radius and $f(Q)$ is the magnetic form factor of the magnetic species used considering Ta$^{4+}$ in our case[26], $g^2 S(S+1)$ is the squared magnetic moment and $Z_n$ and $R_n$ are the coordination number and radial distance of the $n$th nearest-neighbor shell, respectively. For a paramagnetic system without any spin-spin correlations, the $Q$-dependence of magnetic cross section corresponds straightforwardly to the evolution of the absolute squared magnetic moment per atom.

To define reasonable limits for the fitting parameters, we compared our data to those measured on the Lu$_2$Mo$_2$O$_7$ oxide and the Lu$_2$Mo$_2$O$_5$N$_2$ oxynitride[27]. We recall that in our orphan spin scenario the spin moment is expected to be localized in the respective David star consistent with what one would expect in the low-temperature CDW phase. The bump in the oxide data taken after ref. 27 was previously modeled by taking into account the nearest- and the next-nearest-neighbor correlations $\langle S_0 \cdot S_1 \rangle = -0.029$ and $\langle S_0 \cdot S_2 \rangle = -0.056$, respectively, proving the presence of the static short-range molybdenum spin correlations (green dashed line in Fig. 2). On the other hand, the oxynitride data follow the $f^2(Q)$ dependence with zero $\langle S_0 \cdot S_i \rangle$ correlations (red dashed line in Fig. 2). Similar fit of our neutron data on 1T-TaS$_2$ produces the effective magnetic moment of 0.11 $\mu_B$/f.u., representing an upper bound for the magnetic moment of the orphan spins.

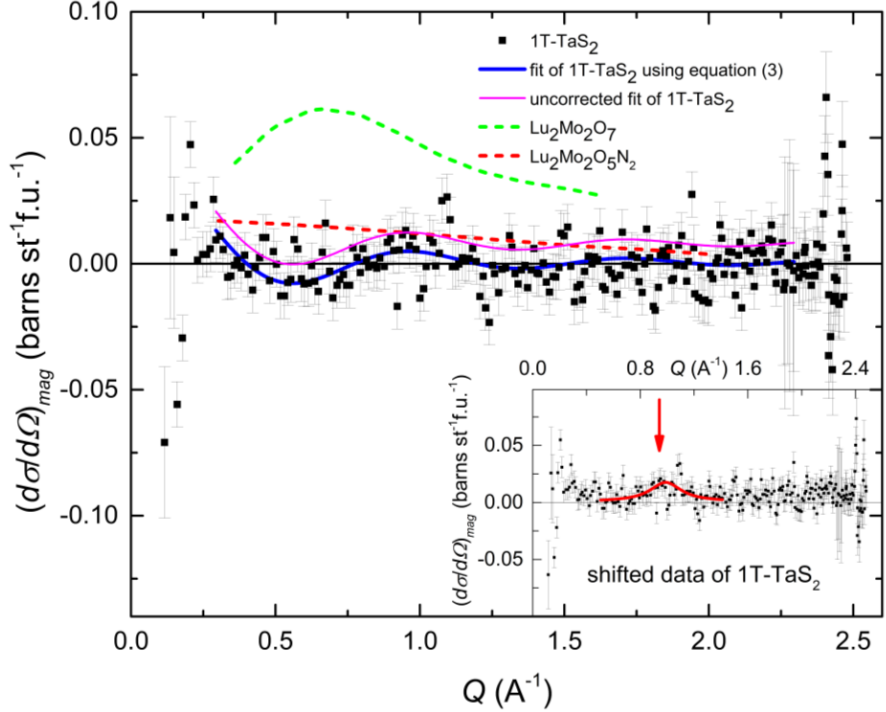

**Fig. 2** The magnetic scattering cross section of 1T-TaS$_2$ at 1.5 K. The inset shows the shifted data (see the text for details) and the Lorentz fit of the bump located at $Q \sim 0.9$ Å$^{-1}$, while the main figure presents the original data together with the fit and the data measured on the Lu$_2$Mo$_2$O$_7$ compound. Vertical error bars represent $1\sigma$ s.d. counting statistics. (Figure 2 shows the data of the Lu$_2$Mo$_2$O$_7$ compound taken from ref. 27 with permission. Copyrighted by the American Physical Society)

As shown in ref. 28, double layers are connected by the $c$-stacking vector across the hexagonal sheets. As a result, the low-temperature phase is commensurate only in the basal plane and not along the $c$-axis because the stacking order established along the $c$-axis is not periodic. Related to this issue, two types of stacking patterns – screw and repetitive - have been previously observed in the ratio ~ 1:2.7 for 1T-TaS$_2$. Due to the large spatial extent of the David star, the nearest neighbors that are expected to contribute to the magnetic cross section are located across the hexagonal layers and not within them. Indeed, assuming that the crucial correlations take place only within the CDW layer, the fit cannot describe satisfactorily the data (shown in Fig. S2 in the Supplemental Material). Therefore, the three-dimensional character of the electronic correlations at low temperatures has to be considered. This result is indeed in good agreement with the recent experimental and theoretical studies, which suggested the importance of interlayer coupling[14, 29-31]. In order to describe the broad peak well by the expression (3), we had to include interactions among the nearest ($r_1 = 5.90$ Å, $Z_1 = 1$) and the next-nearest-neighbor ($r_2 = 8.29$ Å, $Z_2 = 1$) in the repetitive stacking and the next-nearest-neighbor ($r'_2 = 6.79$ Å, $Z'_2 = 1$) in the screw stacking with respect to their ratio, as depicted in Fig. 3.

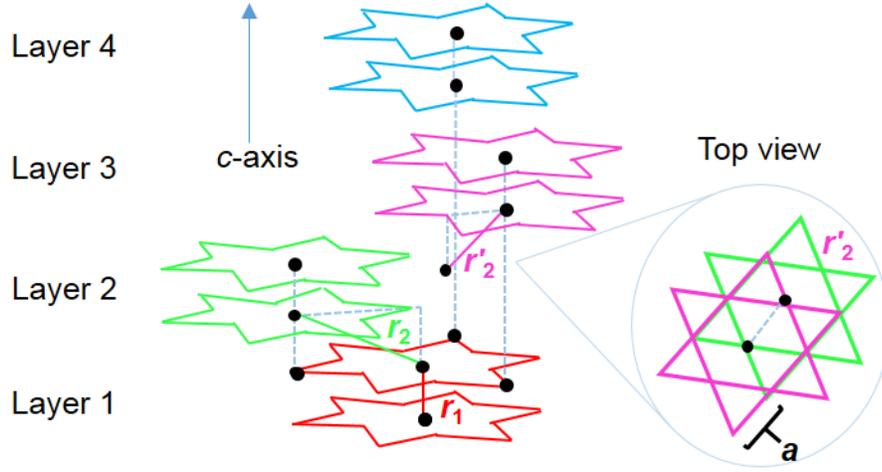

**Fig. 3** CDW structure and interplane stacking in 1T-TaS$_2$. The lattice parameter $a$ = 3.36 Å is depicted in the top view. In the case of the repetitive stacking with the nearest-neighbor distances $r_1$ = 5.90 Å and $r_2$ = 8.29 Å only double layers 1 and 2 alternate while the screw stacking with the nearest-neighbor distances $r_1$ and $r_2'$ = 6.79 Å is formed by double layers 1 to 4.

The best fit to the data by the equation (3) was obtained for the effective magnetic moment of 0.4 ± 0.05 μ$_B$ (≈ 300 % of the expected magnetic moment per each David star) and for the average correlations $\langle S_0 \cdot S_1 \rangle$ = -0.07 ± 0.05, $\langle S_0 \cdot S_2 \rangle$ = 0.17 ± 0.10, $\langle S_0 \cdot S_2 \rangle'$ = 0.39 ± 0.16, respectively (solid blue line in Fig. 2). The solid magenta line represents the fit without the correction on the background term. The magnetic moment is higher compared to the expected $\mu$ ~ 0.13 μ$_B$ value of the magnetic moment per each David star; however, it is necessary to keep in mind that the values of the fitted parameters represent the upper limit. Also, the values and error bars of the average correlations need to be understood in terms of very low signal-to-noise ratio. The dominant correlation is weakly antiferromagnetic, while the second-nearest neighbor correlations are more robust and ferromagnetic; further correlations among more distant David-star patterns can be neglected. Here we note that there is considerable nearest and next-nearest $\langle S_0 \cdot S_i \rangle$ correlations with opposite signs. Interestingly, the ferromagnetic nature of $\langle S_0 \cdot S_i \rangle$ for $i = 2$ corresponds to the positive value of the Curie-Weiss temperatures $\theta_{CW}$ obtained from the magnetization measurement, hinting at subtle ferromagnetic correlations.

In order to study further the short-range magnetic order, we carried out the zero-field (ZF) muon-spin relaxation study from 5 up to 250 K. The ZF spectra can be modeled using the following expressions:

$$G_z(t) = a_i G_{KT}(t) \times G_{mag}(t) + a_{bg}, \quad (4)$$

$$G_{KT}(t) = \frac{1}{3} + \frac{2}{3}(1 - \sigma^2 t^2) exp\left(-\frac{\sigma^2 t^2}{2}\right), \quad (5)$$

$$G_{mag}(t) = exp(-\lambda t), \quad (6)$$

where $a_i$ and $a_{bg}$ are the initial and background asymmetries, respectively. Eq. (5) describes the static Gaussian Kubo-Toyabe function, whereas σ is the nuclear depolarization rate. $G_{mag}(t)$ from Eq. (6) is the magnetic part of the relaxation function, which represents the muon-spin relaxation rate $\lambda$ from the dynamic magnetic fields associated with fluctuating atomic spins. Consistent with the neutron experiment, we observe a sign of the presence of a low-temperature short-ranged magnetic ordering in the asymmetry spectra compared to high temperatures as shown in Fig. 4a and b. The asymmetry spectra can be characterized by the typical Kubo-Toyabe evolution and no pronounced qualitative differences among the spectra are observed for various temperatures (the spectra measured at 5 and 250 K are shown for comparison).

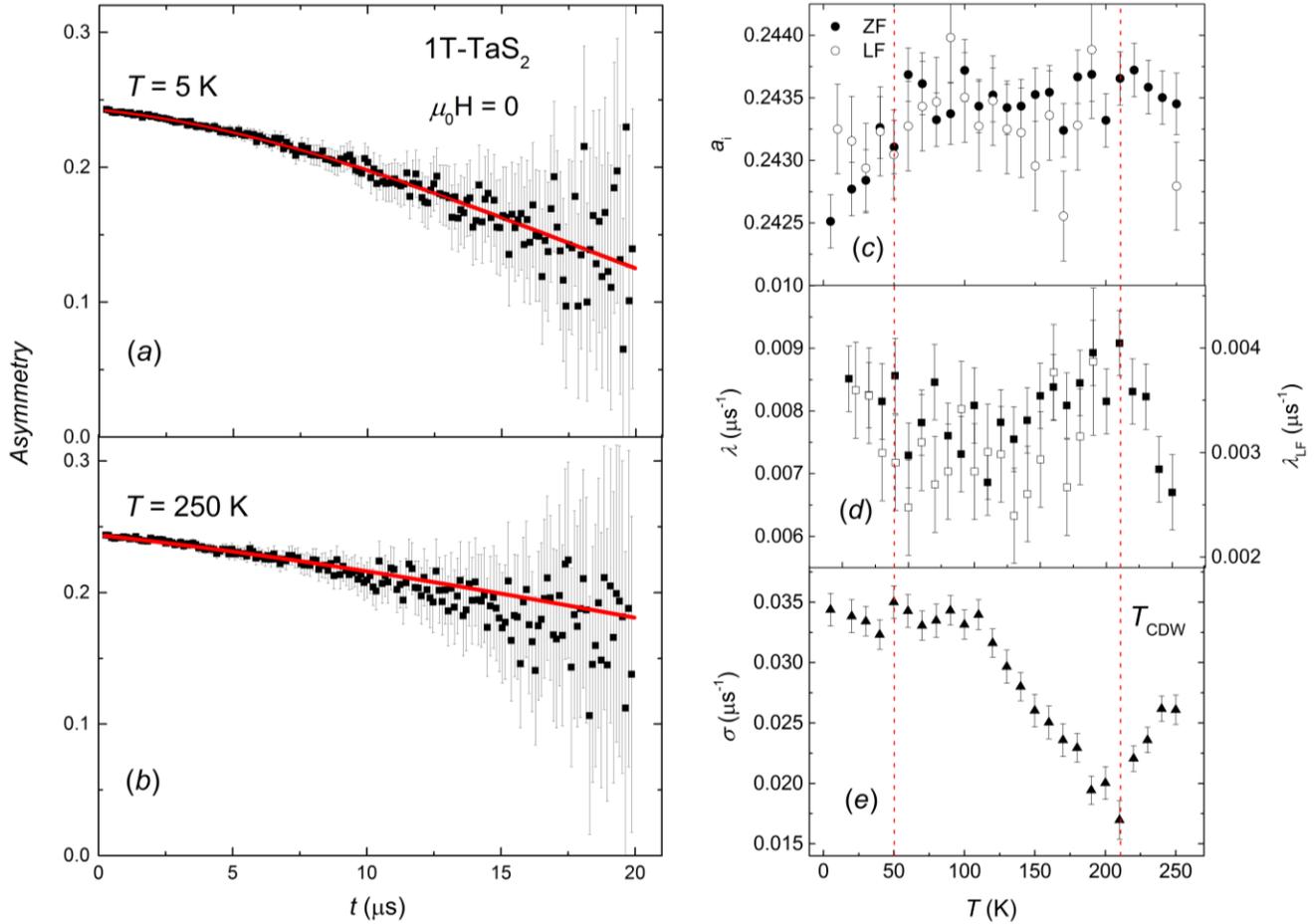

**Fig. 4** The $\mu$SR data of 1T-TaS$_2$. The ZF $\mu$SR asymmetry spectra of 1T-TaS$_2$ are measured at **a** 5 K, and **b** 250 K; the red line represents the fit described in the text. Vertical error bars represent 1$\sigma$ s.d. counting statistics. Temperature dependence of the initial asymmetry both in the ZF and LF mode **c**, the relaxation rate **d**, and the nuclear depolarization rate **e**, taken from the fit described in the text. The vertical error bars for $a_i$, $\lambda$, and $\sigma$ are from fits. The dashed line at 210 K marks the position of the transition into the CDW state upon cooling. The dashed line at ~ 50 K marks the temperature where the trend of the three parameters changes which might be associated with the possible onset of the magnetic short-range order at low temperatures

To suppress the nuclear contribution to the muon asymmetry function, small LF field of 5 mT was applied in a direction parallel to the muon polarization; which is sufficient to fully decouple the muons from the relaxation channel. The effect of the applied longitudinal field also indicates that the magnetism is dynamic in the investigated temperature range (see Fig. S3 in the Supplemental Material). Fig. 4c-e summarizes the results of the Kubo-Toyabe fit over the whole temperature range for the fitting parameters $a_i$, $\lambda$ and $\sigma$. The ZF run of the initial asymmetry parameter shown in Fig. 4c remains temperature-independent down to ~ 50 K, where the asymmetry starts to decrease slightly and at the lowest measured temperature of 5 K. This change points to the onset of a short-range magnetic order. The relaxation rate $\lambda$ in Fig. 4d shows a slight increase below ~ 50 K, implying that this extra relaxation might indicate additional internal magnetic fields. This result by itself is not conclusive as the effect is very weak, in agreement with the recent $\mu$SR measurements performed by Klanjšek *et al.*[19]; however, together with the bulk and neutron data it forms a comprehensive picture about the low-temperature state in 1T-TaS$_2$. The depolarization rate $\sigma$ in Fig. 4e reveals an interesting temperature evolution. Upon warming, it starts to deviate from the constant value at ~ 100 K, which is probably a sign of the muons diffusion, and a minimum of $\sigma$ is observed at ~ 210 K. This value coincides well with the temperature $T_{CDW}$ ~ 210 K (upon warming) of the commensurate CDW transition obtained from the magnetic susceptibility measurement.

## DISCUSSION

With those data from the specific heat and the magnetic susceptibility, we can calculate the Sommerfeld-Wilson ratio. The ratio is given as

$$R_w = \frac{4\pi^2 k_B^2}{3(g\mu B)^2} \frac{\chi_0}{\gamma}. \qquad (7)$$

With $\chi_0 = 5 \cdot 10^{-4}$ emu.mol$^{-1}$ as obtained at the lowest measured temperature $T = 1.8$ K and for $\gamma = 1.84$ mJ.mol$^{-1}$·K$^{-2}$ using the identical sample #4, we get $R_W$ ~ 2 ($R_W = 1$ for the free electron gas) indicating the presence of considerable electronic correlations at low temperatures.

Another remarkable feature in $\chi(T)$ is the absolute size of the jumps $\Delta\chi_1$ and $\Delta\chi_2$ that is consistently seen for all the measured samples (see Table S1 in the Supplemental Material) although the absolute values of $\chi(T)$ are strongly sample-dependent due to different susceptibility term $\chi_0$. From the ratio of the jumps it is possible for us to estimate the relative ratio of the CDW-induced energy gaps as

$$\frac{\Delta\chi_1}{\Delta\chi_2} = \frac{\Delta N(T_{CDW})}{\Delta N(T_{nCDW})} \sim 1.5, \qquad (8)$$

where $\Delta N$ is the electron density of states. The early tunneling experiments on 1T-TaS$_2$[32, 33] were carried out to investigate the size of the charge gaps directly and reported somewhat a larger ratio of ~ 2.5 for charge gap of ~ 10 eV at $T_{CDW}$ and 0.4 eV at $T_{nCDW}$. Later, it was shown that the gap opening at the Fermi level is related to the Mott localization in the commensurate CDW phase[34]. Turning to the question on the low-temperature phase, both the magnetic susceptibility and the specific heat suggest the presence of a correlated phase; another support for this scenario can be found in the neutron data that show a clear bump at a finite Q value. The position of the bump in the diffuse magnetic scattering of 1T-TaS$_2$ is $Q$ ~ 0.9 Å$^{-1}$, close to $2\pi/r_2'$, where $r_2' = 6.79$ Å is the next nearest-neighbor distance across the screw stacking layers. As the respective correlation for this neighbor distance is positive, the $Q$-position implies ferromagnetic correlations between pairs of spins across the hexagonal layers. The correlation length extracted from the full width at half maximum (~ 0.285 ± 0.060 Å$^{-1}$) of the bump (see the Lorentz fit shown in the inset of Fig. 2) is $L$ ~ 22 Å, which roughly corresponds to $4r_1$. Thus, we conclude that correlated domains show a large spatial extent almost over two David-star patterns and across four hexagonal sheets.

Finally, the temperature evolution of the parameters $a_i$ and $\lambda$ obtained by $\mu$SR measurement reveals a change of slope below ~ 50 K. This temperature scale seems to have a physical significance as it corresponds to the abrupt changes in the temperature evolution of specific heat and magnetic susceptibility. Interestingly, a spontaneous ordering into the hidden state in 1T-TaS$_2$ was observed below 100 K after an application of the sub-35-fs laser pulse under non-equilibrium conditions[35]. Our $\mu$SR experiment shows a subtle change in the temperature evolution of the relaxation rate and initial asymmetry value below 50 K, which can be taken as a signature of a low-temperature correlated phase in accordance with the neutron and bulk measurements.

To sum up, we have performed complementary specific heat, magnetization, neutron diffraction, and $\mu$SR measurements of 1T-TaS$_2$ to study the nature of the low-temperature hidden order within the commensurate CDW phase. All the results presented here consistently offer evidences of the "orphan spin" scenario with $S = ½$ moments localized in the centers of the David stars, albeit with a somewhat higher value of the effective magnetic moment as compared to the proposed ~ 0.13 $\mu_B$ per David star. Our results support also the important role of the interlayer correlations in the electronic structure of 1T-TaS$_2$ in line with theoretical studies[30].

## METHODS

Polycrystals and single crystals of 1T-TaS$_2$ was prepared by the solid-state high-temperature reaction of initial elements and by the chemical vapor transport method, respectively (for details of the synthesis, see the Supplemental Material). The samples were characterized by X-ray diffraction (Bruker XRD D8 Discover for powder and Rigaku XtaLAB P200 for single crystals) and confirmed to have the trigonal crystal structure crystallizing in the *P-3m1* space group (see Fig. S1 in the Supplemental Material). The magnetization and heat capacity measurements were carried out using commercial set-ups: MPMS-3 EverCool (Quantum Design, USA) and PPMS-9 equipped with the low-temperature He3 option (Quantum Design, USA).

Polarized diffuse neutron scattering was performed on D7 at Institut Laue-Langevin (ILL) using 4.2 g of sample placed in a double-wall Al cylinder can and incident energy $E_i$ = 3.5 meV ($\lambda$ = 4.8 Å). The temperature was kept at 1.5 K during the whole experimental time. The instrument background was estimated by adding the scattering from an empty and a cadmium-filled sample holder, weighted by the sample transmission. This was subtracted from the data. The data were then subsequently corrected for polarization efficiency using the scattering from amorphous quartz. Vanadium was measured to estimate detector efficiency and to put the data on an absolute scale. Equal time was spent on measuring the scattering along the X, Y and Z directions. The nuclear, magnetic and nuclear spin-incoherent scattering cross sections were separated as a function of both momentum and energy transfer using the 6-point 'xyz'-polarization analysis[36]. The non-spin-flip and spin-flip scattering along each direction was measured with a time ratio of 1:8.

$\mu$SR spectra were measured using the EMU spectrometer at the ISIS Facility, Rutherford Appleton Laboratory, U.K. in the temperature range from 5 to 250 K using a Close Cycle Refrigerator. The sample of the mass of ~ 20 g was enclosed in an Al sample holder. Any exposed Al to the beam was covered by a silver mask (dia. 30 mm) with a 30 μm mylar window for the muons to pass through before entering the sample. The decay-positron asymmetry function was measured as a function of time, in the zero-field (ZF) regime to examine the unperturbed magnetic behavior and in the longitudinal field (LF) of 5 mT.

Data availability

The data that support the findings of this study are available from the corresponding author upon reasonable request.

## ACKNOWLEDGEMENTS

We acknowledge Y. W. Son for useful comments. This work was supported by the Institute for Basic Science (IBS) in Korea (IBS-R009-G1). The experiments were partially performed in MGML (https://mgml.eu/) through the program of Czech Research Infrastructures (LM2011025). We acknowledge the support and beam time at Institut Laue Langevin and ISIS Neutron and Muon Source, in providing neutron and muon research facilities used in this work.

## AUTHOR CONTRIBUTIONS

JGP conceived the idea and supervised the project. MK carried out all the experiments: the neuron scattering experiments were done together with AW while muon experiments were done with AH. LW & SWC prepared the sample. MK and JGP wrote the manuscript with contributions from all the other authors.

## ADDITIONAL INFORMATION

**Supplementary Information** accompanies the paper on the *npj Quantum Materials* website

**Competing interests:** The authors declare that they have no competing financial interests.

**Publisher's note**: Springer Nature remains neutral with regard to jurisdictional claims in published maps and institutional affiliations.